\documentclass[aps,pra,twocolumn,superscriptaddress]{revtex4-2}
\usepackage{amsmath,amssymb,graphicx,bbold}
\usepackage{graphicx}  % needed for figures
\usepackage{dcolumn}   % needed for some tables
\usepackage{bm}        % for math
\usepackage{color}
\usepackage[dvipsnames]{xcolor}
\usepackage{physics}
\usepackage{ulem}
\usepackage{float}
\usepackage{subfigure}

\begin{document}

\title{Observation of triangular-lattice pattern in nonlinear wave mixing with optical vortices}

\author{B. Pinheiro da Silva}
%\email{braianps@gmail.com}
\affiliation{Instituto de F\'{i}sica, Universidade Federal Fluminense, 24210-346 Niter\'{o}i, RJ, Brazil}
\author{G. H. dos Santos}
\affiliation{Departamento de F\'{i}sica, Universidade Federal de Santa Catarina, CEP 88040-900, Florian\'{o}plis, SC, Brazil}
\author{A. G. de Oliveira}
\affiliation{Departamento de F\'{i}sica, Universidade Federal de Santa Catarina, CEP 88040-900, Florian\'{o}plis, SC, Brazil}
\author{N. Rubiano da Silva}
\affiliation{Departamento de F\'{i}sica, Universidade Federal de Santa Catarina, CEP 88040-900, Florian\'{o}plis, SC, Brazil}
\author{W. T. Buono}
\affiliation{School of Physics, University of the Witwatersrand, Private Bag 3, Johannesburg 2050, South Africa}
\author{R. M. Gomes}
\affiliation{Instituto de F\'{i}sica, Universidade Federal de Goi\'as, CEP  74690-900, Goi\^{a}nia, GO, Brazil}
\author{W. C. Soares}
%\email{willamys@fis.ufal.br}
\affiliation{N\'ucleo de Ci\^encias Exatas – NCEx, Universidade Federal de Alagoas, CEP 57309-005, Arapiraca, AL, Brazil}
\author{A. J. Jesus-Silva}
%\email{alcenisio@fis.ufal.br}
\affiliation{Instituto de F\'{i}sica, Universidade Federal de Alagoas, CEP 57072-970, Macei\'{o}, AL, Brazil}
\author{E. J. S. Fonseca}
%\email{eduardo@fis.ufal.br}
\affiliation{Instituto de F\'{i}sica, Universidade Federal de Alagoas, CEP 57072-970, Macei\'{o}, AL, Brazil}
\author{P. H. Souto Ribeiro}
%\email{p.h.s.ribeiro@ufsc.br}
\affiliation{Departamento de F\'{i}sica, Universidade Federal de Santa Catarina, CEP 88040-900, Florian\'{o}plis, SC, Brazil}
\author{A. Z. Khoury}
%\email{azkhoury@id.uff.br}
\affiliation{Instituto de F\'{i}sica, Universidade Federal Fluminense, 24210-346 Niter\'{o}i, RJ, Brazil}
\date{\today}
\begin{abstract}
A triangular-lattice pattern is observed in light beams resulting from the spatial cross modulation between an optical vortex and a triangular shaped beam undergoing parametric interaction. Both up- and down-conversion processes are investigated, and the far-field image of the converted beam exhibits a triangular lattice. The number of sites and the lattice orientation are determined by the topological charge of the vortex beam. In the down-conversion process, the lattice orientation can also be affected by phase conjugation. The observed cross modulation works for a large variety of spatial field structures, and could replace solid-state devices at wavelengths where they are not yet available.
\end{abstract}
%\pacs{03.65.Vf, 03.67.Mn, 42.50.Dv}
%\vskip2pc 
\maketitle

\section{Introduction}
The cross-talk between spatial structures in nonlinear wave mixing  is widely relevant both in classical and quantum regimes. The nonlinear optical process of parametric down-conversion has been extensively employed to generate quantum states of light structured in the transverse spatial degrees of freedom \cite{Walborn10}. In the classical regime, the same process can be operated in the stimulated emission mode (StimPDC) \cite{Ou90,Wang90}, providing a convenient platform for the design of quantum optical schemes \cite{liscidini2013,rozema2015,ciampini2019}, and for the study of the interplay between the spatial structures of the interacting light fields in the parametric process \cite{PH2001,Caetano02,Marcelo18,Andre19,Andre20}. In the same way, parametric up-conversion plays an important role in a wide variety of applications in quantum and classical optical schemes, as for instance frequency conversion of squeezed light fields \cite{Vollmer2014,Kerdoncuff2021} and imaging with visible and invisible light \cite{Barh19, Qiu18}. The spatial structure of light beams, including the so-called optical vortex \cite{padgett2017}, gives rise to interesting effects in up-conversion \cite{dholakia1996,Diney2006,Zhang2010,Hong2014,Liu2018}. Therefore, frequency conversion of structured light paves the way for an increasing number of applications  \cite{Stein2016,wu22}.

In the present work, we investigate the fields generated in the process of parametric up-conversion and stimulated down-conversion. It is known that the nonlinear evolution of optical vortex beams undergoing parametric up- and down-conversion is subjected to selection rules, which determine orbital angular momentum (OAM) conservation as a ubiquitous condition \cite{dholakia1996,Caetano02}, and the appearance of radial modes as a possible side effect depending on the relative chirality of the interacting beams \cite{buono14,buono17,buono18,buono20,Andre21}. Both conditions naturally appear from the straightforward calculation of the spatial overlap between the interacting modes. However, a more appealing physical picture is  to consider the propagation properties of the outgoing field as a result of the spatial cross modulation  due to the nonlinear interaction between the incoming beams, which is equivalent to diffraction through an aperture.

Exploiting this simple physical picture, we demonstrate the occurrence of one 
striking effect in the diffraction phenomena of vortex beams generated in the nonlinear optical process, namely the formation of a triangular lattice in the far-field patterns \cite{Will10,Will18,Shen18}. We observe this outcome both in frequency up- and down- conversion, by mixing a vortex beam with a triangular shaped beam. The triangular lattice in the converted field evinces the effect, and the lattice orientation and number of sites are determined by the topological charge of the incoming vortex beam. In the down-conversion process, the lattice orientation is also affected by phase conjugation \cite{PH2001,Oliveira19}, depending on whether the vortex structure is prepared in the pump or seed beam. Our findings advance the understanding of the role of spatial transverse structures in light fields generated from interaction in a nonlinear medium. Moreover, the fact that these fields have different wavelengths for pump and seed allows wavefront manipulation and sensing in frequency ranges for which there is no commercial modulation devices. 

\section{Spatial cross-modulation in nonlinear wave mixing}
\begin{figure*}[t!]
	\includegraphics[width=1\textwidth]{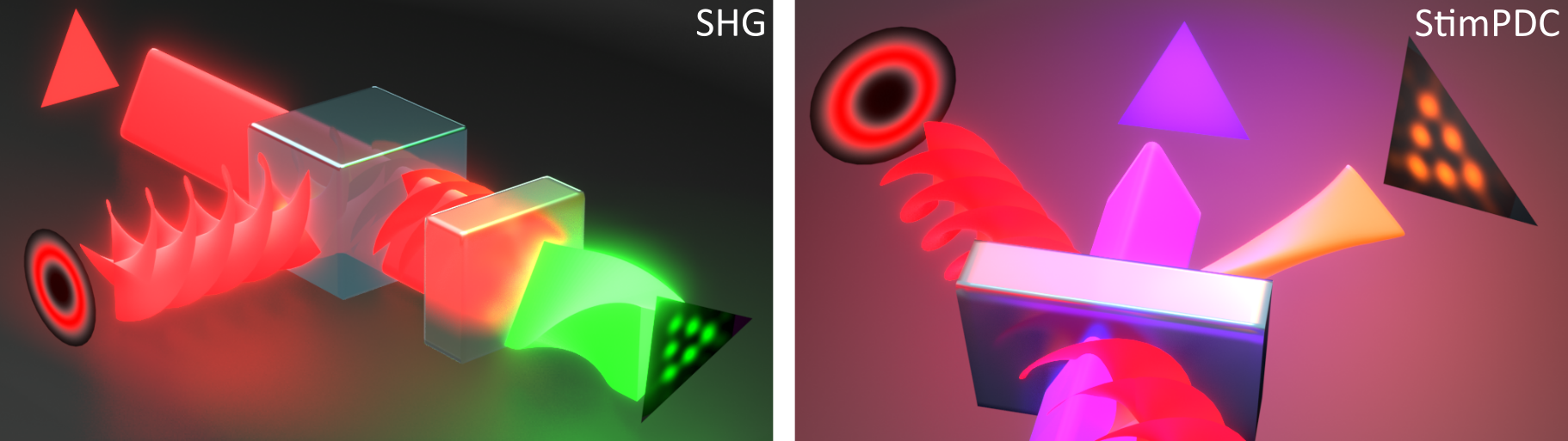}
	\caption{Cross modulation of light fields in nonlinear wave mixing. Left panel: input fields (in red) with orthogonal polarizations, equal frequencies and different spatial structures incident on a polarizing beam splitter for second-harmonic generation (SHG). Right panel: input fields  of different frequencies and spatial structures (in red and purple) in stimulated parametric down-conversion (StimPDC).}
	\label{fig:conv-teo}
\end{figure*}

The wave mixing of two input signals inside a nonlinear crystal generates a new field contribution, 
which is coherently amplified along the interaction length, provided the phase matching condition is 
fulfilled. The phase matching implies a constraint between the wave vectors of the interacting fields \cite{Zhang2017}. 
In the paraxial regime, it is useful to analyse this constraint separately in the longitudinal 
and transverse directions. In the case of an optically thin nonlinear medium, the bandwidth of the 
longitudinal phase matching is large and the spatial profile of the field generated in the 
nonlinear interaction is essentially determined by the product of the transverse structures carried 
by the input beams \cite{bloembergen1977}. The output field carries the combined information of the input beams, in a situation that is quite equivalent to usual diffraction problems, where the field distribution immediately after an obstacle 
is the product between the incident field distribution and the transmission function 
$\mathcal{T}(\mathbf{r})$ that characterizes the obstacle: 
$\mathcal{E}_{out} (\mathbf{r}) = \mathcal{T} (\mathbf{r})\,\mathcal{E}_{in} (\mathbf{r})\,$.
Therefore, the patterns generated in nonlinear wave mixing can be viewed as an effective 
diffraction problem where one input beam plays the role of an obstacle or spatial modulator. 
We next analyse the up- and down-conversion processes separately, demonstrating the striking 
triangular pattern formed by transmission of an optical vortex through a triangular aperture.

% \subsection{Up-conversion}
%

{\it Up-conversion} - In the up-conversion configuration, two input beams $\mathbf{E}_1$ and $\mathbf{E}_2$ are mixed in 
the nonlinear medium and generate the output field $\mathbf{E}_3\,$, satisfying energy 
$\omega_1 + \omega_2 = \omega_3$ and momentum $\mathbf{k}_1 + \mathbf{k}_2 = \mathbf{k}_3$ 
conservation. Each field component carries a spatial structure $\mathcal{E}_j (\mathbf{r})$ 
($j=1,2,3$) and a polarization unit vector $\mathbf{\hat{e}}_j\,$, so that
\begin{equation}
\mathbf{E}_j = \mathcal{E}_j (\mathbf{r})\,\mathbf{\hat{e}}_j\,.
\end{equation}
The spatial structure of the up-converted beam is proportional to the product of those carried by the incoming 
beams \cite{buono18,buono20} 
\begin{equation}
\mathcal{E}_3 (\mathbf{r}) = g\,\mathcal{E}_1 (\mathbf{r})\,\mathcal{E}_2 (\mathbf{r})\,,
\label{eqshg}
\end{equation}
where $g$ is the effective coupling constant. Therefore, the pattern formed by the up-converted beam after the interaction region corresponds to the cross modulation between the input (usually infrared) beams. In this sense, the resulting pattern can be viewed as the diffraction of one beam through an effective transmission function embodied by the other. This interpretation is illustrated in Fig. \ref{fig:conv-teo} (left panel).

% \subsection{Stimulated parametric down-conversion}

{\it Stimulated parametric down-conversion} - In the stimulated down-conversion configuration (StimPDC), two input beams $\mathbf{E}_p$ (pump) 
and $\mathbf{E}_s$ (seed) are mixed in the nonlinear medium and generate the output field 
$\mathbf{E}_i$ (idler), satisfying energy $\omega_p - \omega_s = \omega_i$ and momentum 
$\mathbf{k}_p - \mathbf{k}_s = \mathbf{k}_i$ conservation. Each field component has a spatial 
structure $\mathcal{E}_j (\mathbf{r})$ ($j=p,s,i$) and a polarization unit vector 
$\mathbf{\hat{e}}_j\,$, as before.
The spatial structure of the down-converted beam is proportional to the product between the structure 
carried by the pump and the conjugate of the one carried by the seed beam \cite{PH2001}
\begin{equation}
\mathcal{E}_i (\mathbf{r}) = g\,\mathcal{E}_p (\mathbf{r})\,\mathcal{E}_s^* (\mathbf{r})\,,
\label{eqspdc}
\end{equation}
where $g$ is the effective coupling constant. Therefore, the pattern formed by the down-converted beam after 
the interaction region corresponds to the cross modulation between the pump and the conjugate seed structures. 
In this case, the role of effective transmission function is played differently by the pump and seed beams.
Figure \ref{fig:conv-teo} (right panel) illustrates the situation of having the triangular aperture in the pump field.

\section{Experiment}
\begin{figure}[t!]
        \includegraphics[width=1\columnwidth]{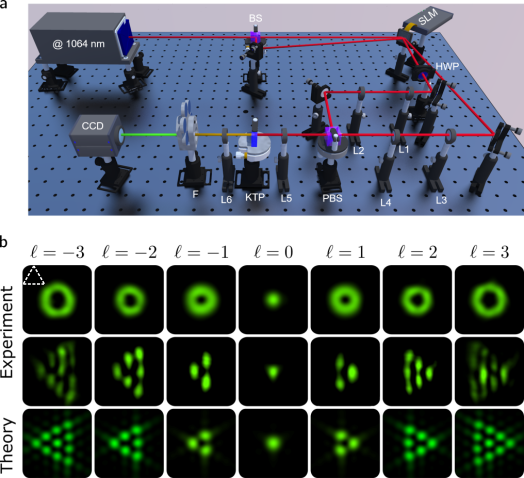}
    \caption{(a) Experimental scheme for spatial cross modulation in up-conversion. BS is beam splitter, SLM is spatial light modulator, HWP is half-waveplate, L1 to L6 are lenses, PBS is polarizing beam splitter, KTP is Potassium Titanyl Phosphate nonlinear crystal, F is a bandpass filter, and CCD is a camera. The power ratio between triangle/LG beam is ~ 1. (b) Measured far-field intensity patterns for the LG input fields (top row) and for the up-converted ones (central row). The bottom row shows the theoretical up-converted patterns. The dashed white triangle illustrates the orientation of the triangular beam.}
    \label{fig:shg-exp}
\end{figure}

{\it Up-conversion setup} - We start by describing the experiment of sum-frequency generation. The experimental setup is sketched in Fig. \ref{fig:shg-exp}a. The horizontally polarized Gaussian beam produced by a 100mW, c.w. Nd:YAG laser ($\lambda = 1064$ nm), which is split in a beam splitter (BS).  One spatial light modulator (SLM) divided in two panels is used to produce a triangular-shaped beam, which is transmitted, and also a Laguerre-Gaussian (LG) mode that is reflected by the BS. In both cases we use the standard modulation approach based on blazed phase gratings and forked masks for the LG modes.  To preserve the transverse  structure along the propagation to the nonlinear crystal, we use  4$f$ imaging lens systems ($f=10$ cm) L1/L2 and L3/L4 in upper and lower paths respectively. The polarization for proper phase matching is set by a half-waveplate oriented at $45^\circ$ in the path of the triangular beam, resulting in vertical polarization. The beams are focused on a potassium titanyl phosphate (KTP) crystal cut for type-II phase matching using a $10$ cm focal length lens. A bandpass filter is used to prevent the non-converted infrared beams to reach the CCD camera, while the up-converted green light ($\lambda = 532$ nm) is imaged after collimation by a 10 cm focal length lens. Far-field intensity profiles are registered with the camera.

Figure \ref{fig:shg-exp}b displays the experimental results for Laguerre-Gaussian (LG) input beams having topological charges ranging from -3 to +3. The images in the upper row show the measured LG-beam intensity profiles. In the second and bottom rows, the measured and theoretical far-field intensity patterns for the up-converted beam are respectively shown.

\begin{figure}[t!]
\includegraphics[width=1\columnwidth]{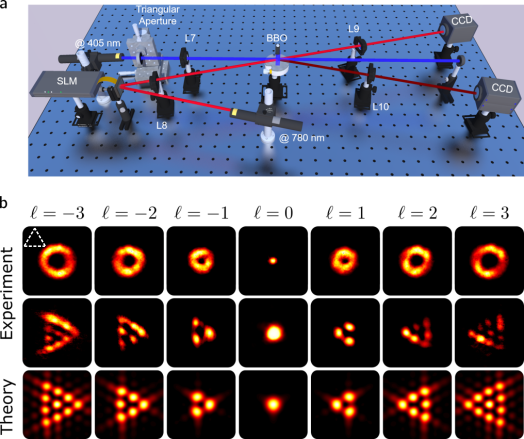}
    \caption{(a) Experimental scheme for spatial cross modulation in stimulated down-conversion.
    SLM is spatial light modulator, L7 to L10 are lenses, BBO is Beta Barium Borate nonlinear crystal, CCD is a camera. The power ratio between triangle/LG beam is ~2. (b) Measured far-field intensity patterns for LG seed (top row) and idler (central row) fields. The bottom row shows the theoretical idler patterns. The dashed white triangle illustrates the orientation of the triangular beam.}
    \label{fig:spdc-exp}
\end{figure}

{\it StimPDC setup} - We have also investigated the StimPDC process. The sketch of the experimental setup is shown in Fig. \ref{fig:spdc-exp}a. We use a vertically polarized, 30 mW, c.w. $405$ nm laser beam, in order to pump a beta barium borate (BBO) nonlinear crystal. The beam is transmitted through a {\em mechanical (not SLM)} triangular aperture, and then imaged in  the crystal plane using a $30$ cm focal length lens. As the seed beam, we use laser light of $780$ nm wavelength and horizontal polarization. We use a SLM to shape the seed beam as LG modes, and the SLM plane is imaged onto the crystal plane using a 30 cm focal length-lens. The pump beam is incident nearly perpendicular to the BBO crystal surface, while the seed beam is incident at about $4$ degrees with respect it. The far-field intensity distributions of both seed and idler beams are registered by CCD cameras with the aid of $40$ cm focal length lenses.

Similarly to the up-conversion measurements, we used LG seed beams having topological charges ranging from -3 to +3. The results are shown in Fig. \ref{fig:spdc-exp}b. The top row displays the measured LG-beams intensity profiles. The images of the measured and theoretically calculated far-field intensity patterns of the idler beam are shown in the intermediate and bottom rows, respectively.

\section{Discussion}

Figures \ref{fig:shg-exp}b and \ref{fig:spdc-exp}b demonstrate the good agreement between experimental and theoretical intensity patterns.  For each conversion process alone, we observe the formation of a triangular lattice with topological charge  $|\ell| = N - 1$, where N is the number of high intensity lobes at the edges, and orientation dependent on the sign of $\ell$. These results reinforce the physical picture presented above, since they follow what is observed when diffracting a LG beam through a triangular aperture \cite{Will10}.

The opposite orientations of the triangular lattices for up-conversion and StimPDC emphasize the phase conjugation effect existing only in StimPDC. Equation~\ref{eqspdc} shows the dependence of the idler field on the phase conjugated seed field $\mathcal{E}_s^* (\mathbf{r})$. In this case, because the seed is prepared as a LG beam with topological charge $+\ell$, the triangular lattice formed in the idler looks like it was the diffraction of a $-\ell$ beam. 

The results also show that the effective spatial modulation in nonlinear wave mixing is of both amplitude and phase. Even though the phase modulation effects is more clearly demonstrated for StimPDC, it also works for up-conversion.

\section{Conclusion}

In conclusion, we demonstrated triangular-lattice patterns generated by nonlinear wave mixing of an optical vortex with a triangular aperture-shaped beam, which works as a spatial modulation device. The cross modulation between input optical fields in the conversion schemes is, however, more general, and could be used to overcome the lack of devices in certain frequency ranges, whereas its counterpart in the visible range is readily available. Wavefront shaping in the THz, in the extreme ultraviolet and in the x-ray ranges, for instance, can be achieved by using a spatial light modulator to control the visible input field. In the THz range, for instance, nonlinear optical conversion from visible light is already used to generate \cite{bai2019} and detect \cite{haase2019} THz fields, and a scheme to optically control metasurfaces generating THz radiation has been recently demonstrated \cite{jana2021}. In addition,  wavefront sensing of telecom and x-ray fields can be accomplished by conversion to the visible range. The phase information thus transferred to the visible output field can be recorded using a common CCD camera. One example of such application is the conversion of infrared images to visibile \cite{Qiu18} using up-conversion. In the up-conversion experiment of our work we use the same kind of cross modulation for a different purpose, to demonstrate detection of topological charges.

Moreover, our scheme of StimPDC allows for filtering of phase information. Recently Rocha et al. \cite{rocha2021} introduced a way of filtering the random phase from speckles through nonlinear wave mixing. The configuration presented there works only if the conjugate phase is present in one of patterns; a restriction that is naturally lifted in StimPDC. A possible application would be using a random phase or medium as an encryption key in optical communication. An image (seed beam) encodes information, which is transferred to the idler beam in StimPDC. The decoded information would be obtained by propagating the idler through the key. Directly filtering the random phase in a speckle pattern is of paramount importance also to imaging systems, and mode sorters, where the information is commonly achieved through computationally extensive post-processing using statistical correlation.

Our findings advance the knowledge about the role of spatially structured light in nonlinear wave mixing. The theoretical modelling and experimental control of frequency conversion processes is crucial in applications like quantum communication, and quantum memories and relays \cite{Castelvecchi21}.

\begin{acknowledgements}
The authors would like to thank the Brazilian Agencies, Conselho Nacional de Desenvolvimento Tecnol\'ogico (CNPq), Funda\c c\~{a}o Carlos Chagas Filho de Amparo \`{a} Pesquisa do Estado do Rio de Janeiro (FAPERJ), Funda\c c\~{a}o de Amparo \`{a} Pesquisa e Inova\c{c}\~{a}o do Estado de Santa Catarina (FAPESC), Funda\c c\~{a}o de Amparo \`{a} Pesquisa e Inova\c{c}\~{a}o do Estado de Goi\'{a}s
(FAPEG) and the Brazilian National Institute of Science and Technology of Quantum Information (INCT/IQ). This study was funded in part by the Coordena\c c\~{a}o de Aperfei\c coamento de Pessoal de N\'ivel Superior - Brasil (CAPES) - Finance Code 001.
\end{acknowledgements}

\bibliography{bibfile}

\begin{thebibliography}{39}%
\makeatletter
\providecommand \@ifxundefined [1]{%
 \@ifx{#1\undefined}
}%
\providecommand \@ifnum [1]{%
 \ifnum #1\expandafter \@firstoftwo
 \else \expandafter \@secondoftwo
 \fi
}%
\providecommand \@ifx [1]{%
 \ifx #1\expandafter \@firstoftwo
 \else \expandafter \@secondoftwo
 \fi
}%
\providecommand \natexlab [1]{#1}%
\providecommand \enquote  [1]{``#1''}%
\providecommand \bibnamefont  [1]{#1}%
\providecommand \bibfnamefont [1]{#1}%
\providecommand \citenamefont [1]{#1}%
\providecommand \href@noop [0]{\@secondoftwo}%
\providecommand \href [0]{\begingroup \@sanitize@url \@href}%
\providecommand \@href[1]{\@@startlink{#1}\@@href}%
\providecommand \@@href[1]{\endgroup#1\@@endlink}%
\providecommand \@sanitize@url [0]{\catcode `\\12\catcode `\$12\catcode
  `\&12\catcode `\#12\catcode `\^12\catcode `\_12\catcode `\%12\relax}%
\providecommand \@@startlink[1]{}%
\providecommand \@@endlink[0]{}%
\providecommand \url  [0]{\begingroup\@sanitize@url \@url }%
\providecommand \@url [1]{\endgroup\@href {#1}{\urlprefix }}%
\providecommand \urlprefix  [0]{URL }%
\providecommand \Eprint [0]{\href }%
\providecommand \doibase [0]{https://doi.org/}%
\providecommand \selectlanguage [0]{\@gobble}%
\providecommand \bibinfo  [0]{\@secondoftwo}%
\providecommand \bibfield  [0]{\@secondoftwo}%
\providecommand \translation [1]{[#1]}%
\providecommand \BibitemOpen [0]{}%
\providecommand \bibitemStop [0]{}%
\providecommand \bibitemNoStop [0]{.\EOS\space}%
\providecommand \EOS [0]{\spacefactor3000\relax}%
\providecommand \BibitemShut  [1]{\csname bibitem#1\endcsname}%
\let\auto@bib@innerbib\@empty
%</preamble>
\bibitem [{\citenamefont {Walborn}\ \emph {et~al.}(2010)\citenamefont
  {Walborn}, \citenamefont {Monken}, \citenamefont {Pádua},\ and\
  \citenamefont {{Souto Ribeiro}}}]{Walborn10}%
  \BibitemOpen
  \bibfield  {author} {\bibinfo {author} {\bibfnamefont {S.}~\bibnamefont
  {Walborn}}, \bibinfo {author} {\bibfnamefont {C.}~\bibnamefont {Monken}},
  \bibinfo {author} {\bibfnamefont {S.}~\bibnamefont {Pádua}},\ and\ \bibinfo
  {author} {\bibfnamefont {P.}~\bibnamefont {{Souto Ribeiro}}},\ }\bibfield
  {title} {\bibinfo {title} {Spatial correlations in parametric
  down-conversion},\ }\href
  {https://doi.org/https://doi.org/10.1016/j.physrep.2010.06.003} {\bibfield
  {journal} {\bibinfo  {journal} {Physics Reports}\ }\textbf {\bibinfo {volume}
  {495}},\ \bibinfo {pages} {87} (\bibinfo {year} {2010})}\BibitemShut
  {NoStop}%
\bibitem [{\citenamefont {Ou}\ \emph {et~al.}(1990{\natexlab{a}})\citenamefont
  {Ou}, \citenamefont {Wang},\ and\ \citenamefont {Mandel}}]{Ou90}%
  \BibitemOpen
  \bibfield  {author} {\bibinfo {author} {\bibfnamefont {Z.~Y.}\ \bibnamefont
  {Ou}}, \bibinfo {author} {\bibfnamefont {L.~J.}\ \bibnamefont {Wang}},\ and\
  \bibinfo {author} {\bibfnamefont {L.}~\bibnamefont {Mandel}},\ }\bibfield
  {title} {\bibinfo {title} {Photon amplification by parametric
  downconversion},\ }\href {https://doi.org/10.1364/JOSAB.7.000211} {\bibfield
  {journal} {\bibinfo  {journal} {J. Opt. Soc. Am. B}\ }\textbf {\bibinfo
  {volume} {7}},\ \bibinfo {pages} {211} (\bibinfo {year}
  {1990}{\natexlab{a}})}\BibitemShut {NoStop}%
\bibitem [{\citenamefont {Ou}\ \emph {et~al.}(1990{\natexlab{b}})\citenamefont
  {Ou}, \citenamefont {Wang}, \citenamefont {Zou},\ and\ \citenamefont
  {Mandel}}]{Wang90}%
  \BibitemOpen
  \bibfield  {author} {\bibinfo {author} {\bibfnamefont {Z.~Y.}\ \bibnamefont
  {Ou}}, \bibinfo {author} {\bibfnamefont {L.~J.}\ \bibnamefont {Wang}},
  \bibinfo {author} {\bibfnamefont {X.~Y.}\ \bibnamefont {Zou}},\ and\ \bibinfo
  {author} {\bibfnamefont {L.}~\bibnamefont {Mandel}},\ }\bibfield  {title}
  {\bibinfo {title} {Coherence in two-photon down-conversion induced by a
  laser},\ }\href {https://doi.org/10.1103/PhysRevA.41.1597} {\bibfield
  {journal} {\bibinfo  {journal} {Phys. Rev. A}\ }\textbf {\bibinfo {volume}
  {41}},\ \bibinfo {pages} {1597} (\bibinfo {year}
  {1990}{\natexlab{b}})}\BibitemShut {NoStop}%
\bibitem [{\citenamefont {Liscidini}\ and\ \citenamefont
  {Sipe}(2013)}]{liscidini2013}%
  \BibitemOpen
  \bibfield  {author} {\bibinfo {author} {\bibfnamefont {M.}~\bibnamefont
  {Liscidini}}\ and\ \bibinfo {author} {\bibfnamefont {J.~E.}\ \bibnamefont
  {Sipe}},\ }\bibfield  {title} {\bibinfo {title} {Stimulated {{Emission
  Tomography}}},\ }\href {https://doi.org/10.1103/PhysRevLett.111.193602}
  {\bibfield  {journal} {\bibinfo  {journal} {Phys. Rev. Lett.}\ }\textbf
  {\bibinfo {volume} {111}},\ \bibinfo {pages} {193602} (\bibinfo {year}
  {2013})}\BibitemShut {NoStop}%
\bibitem [{\citenamefont {Rozema}\ \emph {et~al.}(2015)\citenamefont {Rozema},
  \citenamefont {Wang}, \citenamefont {Mahler}, \citenamefont {Hayat},
  \citenamefont {Steinberg}, \citenamefont {Sipe},\ and\ \citenamefont
  {Liscidini}}]{rozema2015}%
  \BibitemOpen
  \bibfield  {author} {\bibinfo {author} {\bibfnamefont {L.~A.}\ \bibnamefont
  {Rozema}}, \bibinfo {author} {\bibfnamefont {C.}~\bibnamefont {Wang}},
  \bibinfo {author} {\bibfnamefont {D.~H.}\ \bibnamefont {Mahler}}, \bibinfo
  {author} {\bibfnamefont {A.}~\bibnamefont {Hayat}}, \bibinfo {author}
  {\bibfnamefont {A.~M.}\ \bibnamefont {Steinberg}}, \bibinfo {author}
  {\bibfnamefont {J.~E.}\ \bibnamefont {Sipe}},\ and\ \bibinfo {author}
  {\bibfnamefont {M.}~\bibnamefont {Liscidini}},\ }\bibfield  {title} {\bibinfo
  {title} {Characterizing an entangled-photon source with classical detectors
  and measurements},\ }\href {https://doi.org/10.1364/OPTICA.2.000430}
  {\bibfield  {journal} {\bibinfo  {journal} {Optica, OPTICA}\ }\textbf
  {\bibinfo {volume} {2}},\ \bibinfo {pages} {430} (\bibinfo {year}
  {2015})}\BibitemShut {NoStop}%
\bibitem [{\citenamefont {Ciampini}\ \emph {et~al.}(2019)\citenamefont
  {Ciampini}, \citenamefont {Geraldi}, \citenamefont {Cimini}, \citenamefont
  {Macchiavello}, \citenamefont {Sipe}, \citenamefont {Liscidini},\ and\
  \citenamefont {Mataloni}}]{ciampini2019}%
  \BibitemOpen
  \bibfield  {author} {\bibinfo {author} {\bibfnamefont {M.~A.}\ \bibnamefont
  {Ciampini}}, \bibinfo {author} {\bibfnamefont {A.}~\bibnamefont {Geraldi}},
  \bibinfo {author} {\bibfnamefont {V.}~\bibnamefont {Cimini}}, \bibinfo
  {author} {\bibfnamefont {C.}~\bibnamefont {Macchiavello}}, \bibinfo {author}
  {\bibfnamefont {J.~E.}\ \bibnamefont {Sipe}}, \bibinfo {author}
  {\bibfnamefont {M.}~\bibnamefont {Liscidini}},\ and\ \bibinfo {author}
  {\bibfnamefont {P.}~\bibnamefont {Mataloni}},\ }\bibfield  {title} {\bibinfo
  {title} {Stimulated emission tomography: Beyond polarization},\ }\href
  {https://doi.org/10.1364/OL.44.000041} {\bibfield  {journal} {\bibinfo
  {journal} {Opt. Lett., OL}\ }\textbf {\bibinfo {volume} {44}},\ \bibinfo
  {pages} {41} (\bibinfo {year} {2019})}\BibitemShut {NoStop}%
\bibitem [{\citenamefont {Souto~Ribeiro}\ \emph {et~al.}(2001)\citenamefont
  {Souto~Ribeiro}, \citenamefont {Caetano}, \citenamefont {Almeida},
  \citenamefont {Huguenin}, \citenamefont {Coutinho~dos Santos},\ and\
  \citenamefont {Khoury}}]{PH2001}%
  \BibitemOpen
  \bibfield  {author} {\bibinfo {author} {\bibfnamefont {P.~H.}\ \bibnamefont
  {Souto~Ribeiro}}, \bibinfo {author} {\bibfnamefont {D.~P.}\ \bibnamefont
  {Caetano}}, \bibinfo {author} {\bibfnamefont {M.~P.}\ \bibnamefont
  {Almeida}}, \bibinfo {author} {\bibfnamefont {J.~A.}\ \bibnamefont
  {Huguenin}}, \bibinfo {author} {\bibfnamefont {B.}~\bibnamefont {Coutinho~dos
  Santos}},\ and\ \bibinfo {author} {\bibfnamefont {A.~Z.}\ \bibnamefont
  {Khoury}},\ }\bibfield  {title} {\bibinfo {title} {Observation of image
  transfer and phase conjugation in stimulated down-conversion},\ }\href
  {https://doi.org/10.1103/PhysRevLett.87.133602} {\bibfield  {journal}
  {\bibinfo  {journal} {Phys. Rev. Lett.}\ }\textbf {\bibinfo {volume} {87}},\
  \bibinfo {pages} {133602} (\bibinfo {year} {2001})}\BibitemShut {NoStop}%
\bibitem [{\citenamefont {Caetano}\ \emph {et~al.}(2002)\citenamefont
  {Caetano}, \citenamefont {Almeida}, \citenamefont {Souto~Ribeiro},
  \citenamefont {Huguenin}, \citenamefont {Coutinho~dos Santos},\ and\
  \citenamefont {Khoury}}]{Caetano02}%
  \BibitemOpen
  \bibfield  {author} {\bibinfo {author} {\bibfnamefont {D.~P.}\ \bibnamefont
  {Caetano}}, \bibinfo {author} {\bibfnamefont {M.~P.}\ \bibnamefont
  {Almeida}}, \bibinfo {author} {\bibfnamefont {P.~H.}\ \bibnamefont
  {Souto~Ribeiro}}, \bibinfo {author} {\bibfnamefont {J.~A.~O.}\ \bibnamefont
  {Huguenin}}, \bibinfo {author} {\bibfnamefont {B.}~\bibnamefont {Coutinho~dos
  Santos}},\ and\ \bibinfo {author} {\bibfnamefont {A.~Z.}\ \bibnamefont
  {Khoury}},\ }\bibfield  {title} {\bibinfo {title} {Conservation of orbital
  angular momentum in stimulated down-conversion},\ }\href
  {https://doi.org/10.1103/PhysRevA.66.041801} {\bibfield  {journal} {\bibinfo
  {journal} {Phys. Rev. A}\ }\textbf {\bibinfo {volume} {66}},\ \bibinfo
  {pages} {041801} (\bibinfo {year} {2002})}\BibitemShut {NoStop}%
\bibitem [{\citenamefont {Arruda}\ \emph {et~al.}(2018)\citenamefont {Arruda},
  \citenamefont {Soares}, \citenamefont {Walborn}, \citenamefont {Tasca},
  \citenamefont {Kanaan}, \citenamefont {Medeiros~de Ara\'ujo},\ and\
  \citenamefont {Souto~Ribeiro}}]{Marcelo18}%
  \BibitemOpen
  \bibfield  {author} {\bibinfo {author} {\bibfnamefont {M.~F.~Z.}\
  \bibnamefont {Arruda}}, \bibinfo {author} {\bibfnamefont {W.~C.}\
  \bibnamefont {Soares}}, \bibinfo {author} {\bibfnamefont {S.~P.}\
  \bibnamefont {Walborn}}, \bibinfo {author} {\bibfnamefont {D.~S.}\
  \bibnamefont {Tasca}}, \bibinfo {author} {\bibfnamefont {A.}~\bibnamefont
  {Kanaan}}, \bibinfo {author} {\bibfnamefont {R.}~\bibnamefont {Medeiros~de
  Ara\'ujo}},\ and\ \bibinfo {author} {\bibfnamefont {P.~H.}\ \bibnamefont
  {Souto~Ribeiro}},\ }\bibfield  {title} {\bibinfo {title} {Klyshko's
  advanced-wave picture in stimulated parametric down-conversion with a
  spatially structured pump beam},\ }\href
  {https://doi.org/10.1103/PhysRevA.98.023850} {\bibfield  {journal} {\bibinfo
  {journal} {Phys. Rev. A}\ }\textbf {\bibinfo {volume} {98}},\ \bibinfo
  {pages} {023850} (\bibinfo {year} {2018})}\BibitemShut {NoStop}%
\bibitem [{\citenamefont {{de Oliveira}}\ \emph
  {et~al.}(2020{\natexlab{a}})\citenamefont {{de Oliveira}}, \citenamefont
  {Arruda}, \citenamefont {Soares}, \citenamefont {Walborn}, \citenamefont
  {Gomes}, \citenamefont {{Medeiros de Ara{\'u}jo}},\ and\ \citenamefont
  {Souto~Ribeiro}}]{Andre19}%
  \BibitemOpen
  \bibfield  {author} {\bibinfo {author} {\bibfnamefont {A.~G.}\ \bibnamefont
  {{de Oliveira}}}, \bibinfo {author} {\bibfnamefont {M.~F.~Z.}\ \bibnamefont
  {Arruda}}, \bibinfo {author} {\bibfnamefont {W.~C.}\ \bibnamefont {Soares}},
  \bibinfo {author} {\bibfnamefont {S.~P.}\ \bibnamefont {Walborn}}, \bibinfo
  {author} {\bibfnamefont {R.~M.}\ \bibnamefont {Gomes}}, \bibinfo {author}
  {\bibfnamefont {R.}~\bibnamefont {{Medeiros de Ara{\'u}jo}}},\ and\ \bibinfo
  {author} {\bibfnamefont {P.~H.}\ \bibnamefont {Souto~Ribeiro}},\ }\bibfield
  {title} {\bibinfo {title} {Real-{{Time Phase Conjugation}} of {{Vector Vortex
  Beams}}},\ }\href {https://doi.org/10.1021/acsphotonics.9b01524} {\bibfield
  {journal} {\bibinfo  {journal} {ACS Photonics}\ }\textbf {\bibinfo {volume}
  {7}},\ \bibinfo {pages} {249} (\bibinfo {year}
  {2020}{\natexlab{a}})}\BibitemShut {NoStop}%
\bibitem [{\citenamefont {{de Oliveira}}\ \emph
  {et~al.}(2020{\natexlab{b}})\citenamefont {{de Oliveira}}, \citenamefont
  {{Rubiano da Silva}}, \citenamefont {{Medeiros de Ara{\'u}jo}}, \citenamefont
  {Souto~Ribeiro},\ and\ \citenamefont {Walborn}}]{Andre20}%
  \BibitemOpen
  \bibfield  {author} {\bibinfo {author} {\bibfnamefont {A.~G.}\ \bibnamefont
  {{de Oliveira}}}, \bibinfo {author} {\bibfnamefont {N.}~\bibnamefont
  {{Rubiano da Silva}}}, \bibinfo {author} {\bibfnamefont {R.}~\bibnamefont
  {{Medeiros de Ara{\'u}jo}}}, \bibinfo {author} {\bibfnamefont {P.~H.}\
  \bibnamefont {Souto~Ribeiro}},\ and\ \bibinfo {author} {\bibfnamefont
  {S.~P.}\ \bibnamefont {Walborn}},\ }\bibfield  {title} {\bibinfo {title}
  {Quantum {{Optical Description}} of {{Phase Conjugation}} of {{Vector Vortex
  Beams}} in {{Stimulated Parametric Down}}-{{Conversion}}},\ }\href
  {https://doi.org/10.1103/PhysRevApplied.14.024048} {\bibfield  {journal}
  {\bibinfo  {journal} {Phys. Rev. Appl.}\ }\textbf {\bibinfo {volume} {14}},\
  \bibinfo {pages} {024048} (\bibinfo {year} {2020}{\natexlab{b}})}\BibitemShut
  {NoStop}%
\bibitem [{\citenamefont {Vollmer}\ \emph {et~al.}(2014)\citenamefont
  {Vollmer}, \citenamefont {Baune}, \citenamefont {Samblowski}, \citenamefont
  {Eberle}, \citenamefont {H\"andchen}, \citenamefont
  {Fiur\'a\ifmmode~\check{s}\else \v{s}\fi{}ek},\ and\ \citenamefont
  {Schnabel}}]{Vollmer2014}%
  \BibitemOpen
  \bibfield  {author} {\bibinfo {author} {\bibfnamefont {C.~E.}\ \bibnamefont
  {Vollmer}}, \bibinfo {author} {\bibfnamefont {C.}~\bibnamefont {Baune}},
  \bibinfo {author} {\bibfnamefont {A.}~\bibnamefont {Samblowski}}, \bibinfo
  {author} {\bibfnamefont {T.}~\bibnamefont {Eberle}}, \bibinfo {author}
  {\bibfnamefont {V.}~\bibnamefont {H\"andchen}}, \bibinfo {author}
  {\bibfnamefont {J.}~\bibnamefont {Fiur\'a\ifmmode~\check{s}\else
  \v{s}\fi{}ek}},\ and\ \bibinfo {author} {\bibfnamefont {R.}~\bibnamefont
  {Schnabel}},\ }\bibfield  {title} {\bibinfo {title} {Quantum up-conversion of
  squeezed vacuum states from 1550 to 532 nm},\ }\href
  {https://doi.org/10.1103/PhysRevLett.112.073602} {\bibfield  {journal}
  {\bibinfo  {journal} {Phys. Rev. Lett.}\ }\textbf {\bibinfo {volume} {112}},\
  \bibinfo {pages} {073602} (\bibinfo {year} {2014})}\BibitemShut {NoStop}%
\bibitem [{\citenamefont {Kerdoncuff}\ \emph {et~al.}(2021)\citenamefont
  {Kerdoncuff}, \citenamefont {Christensen},\ and\ \citenamefont
  {Lassen}}]{Kerdoncuff2021}%
  \BibitemOpen
  \bibfield  {author} {\bibinfo {author} {\bibfnamefont {H.}~\bibnamefont
  {Kerdoncuff}}, \bibinfo {author} {\bibfnamefont {J.~B.}\ \bibnamefont
  {Christensen}},\ and\ \bibinfo {author} {\bibfnamefont {M.}~\bibnamefont
  {Lassen}},\ }\bibfield  {title} {\bibinfo {title} {Quantum frequency
  conversion of vacuum squeezed light to bright tunable blue squeezed light and
  higher-order spatial modes},\ }\href {https://doi.org/10.1364/OE.436325}
  {\bibfield  {journal} {\bibinfo  {journal} {Opt. Express}\ }\textbf {\bibinfo
  {volume} {29}},\ \bibinfo {pages} {29828} (\bibinfo {year}
  {2021})}\BibitemShut {NoStop}%
\bibitem [{\citenamefont {Barh}\ \emph {et~al.}(2019)\citenamefont {Barh},
  \citenamefont {Rodrigo}, \citenamefont {Meng}, \citenamefont {Pedersen},\
  and\ \citenamefont {Tidemand-Lichtenberg}}]{Barh19}%
  \BibitemOpen
  \bibfield  {author} {\bibinfo {author} {\bibfnamefont {A.}~\bibnamefont
  {Barh}}, \bibinfo {author} {\bibfnamefont {P.~J.}\ \bibnamefont {Rodrigo}},
  \bibinfo {author} {\bibfnamefont {L.}~\bibnamefont {Meng}}, \bibinfo {author}
  {\bibfnamefont {C.}~\bibnamefont {Pedersen}},\ and\ \bibinfo {author}
  {\bibfnamefont {P.}~\bibnamefont {Tidemand-Lichtenberg}},\ }\bibfield
  {title} {\bibinfo {title} {Parametric upconversion imaging and its
  applications},\ }\href {https://doi.org/10.1364/AOP.11.000952} {\bibfield
  {journal} {\bibinfo  {journal} {Adv. Opt. Photon.}\ }\textbf {\bibinfo
  {volume} {11}},\ \bibinfo {pages} {952} (\bibinfo {year} {2019})}\BibitemShut
  {NoStop}%
\bibitem [{\citenamefont {Qiu}\ \emph {et~al.}(2018)\citenamefont {Qiu},
  \citenamefont {Li}, \citenamefont {Zhang}, \citenamefont {Zhu},\ and\
  \citenamefont {Chen}}]{Qiu18}%
  \BibitemOpen
  \bibfield  {author} {\bibinfo {author} {\bibfnamefont {X.}~\bibnamefont
  {Qiu}}, \bibinfo {author} {\bibfnamefont {F.}~\bibnamefont {Li}}, \bibinfo
  {author} {\bibfnamefont {W.}~\bibnamefont {Zhang}}, \bibinfo {author}
  {\bibfnamefont {Z.}~\bibnamefont {Zhu}},\ and\ \bibinfo {author}
  {\bibfnamefont {L.}~\bibnamefont {Chen}},\ }\bibfield  {title} {\bibinfo
  {title} {Spiral phase contrast imaging in nonlinear optics: seeing phase
  objects using invisible illumination},\ }\href
  {https://doi.org/10.1364/OPTICA.5.000208} {\bibfield  {journal} {\bibinfo
  {journal} {Optica}\ }\textbf {\bibinfo {volume} {5}},\ \bibinfo {pages} {208}
  (\bibinfo {year} {2018})}\BibitemShut {NoStop}%
\bibitem [{\citenamefont {Padgett}(2017)}]{padgett2017}%
  \BibitemOpen
  \bibfield  {author} {\bibinfo {author} {\bibfnamefont {M.~J.}\ \bibnamefont
  {Padgett}},\ }\bibfield  {title} {\bibinfo {title} {Orbital angular momentum
  25 years on [{{Invited}}]},\ }\href {https://doi.org/10.1364/OE.25.011265}
  {\bibfield  {journal} {\bibinfo  {journal} {Opt. Express, OE}\ }\textbf
  {\bibinfo {volume} {25}},\ \bibinfo {pages} {11265} (\bibinfo {year}
  {2017})}\BibitemShut {NoStop}%
\bibitem [{\citenamefont {Dholakia}\ \emph {et~al.}(1996)\citenamefont
  {Dholakia}, \citenamefont {Simpson}, \citenamefont {Padgett},\ and\
  \citenamefont {Allen}}]{dholakia1996}%
  \BibitemOpen
  \bibfield  {author} {\bibinfo {author} {\bibfnamefont {K.}~\bibnamefont
  {Dholakia}}, \bibinfo {author} {\bibfnamefont {N.~B.}\ \bibnamefont
  {Simpson}}, \bibinfo {author} {\bibfnamefont {M.~J.}\ \bibnamefont
  {Padgett}},\ and\ \bibinfo {author} {\bibfnamefont {L.}~\bibnamefont
  {Allen}},\ }\bibfield  {title} {\bibinfo {title} {Second-harmonic generation
  and the orbital angular momentum of light},\ }\href
  {https://doi.org/10.1103/PhysRevA.54.R3742} {\bibfield  {journal} {\bibinfo
  {journal} {Phys. Rev. A}\ }\textbf {\bibinfo {volume} {54}},\ \bibinfo
  {pages} {R3742} (\bibinfo {year} {1996})}\BibitemShut {NoStop}%
\bibitem [{\citenamefont {Ether}\ \emph {et~al.}(2006)\citenamefont {Ether},
  \citenamefont {Souto~Ribeiro}, \citenamefont {Monken},\ and\ \citenamefont
  {de~Matos~Filho}}]{Diney2006}%
  \BibitemOpen
  \bibfield  {author} {\bibinfo {author} {\bibfnamefont {D.~S.}\ \bibnamefont
  {Ether}}, \bibinfo {author} {\bibfnamefont {P.~H.}\ \bibnamefont
  {Souto~Ribeiro}}, \bibinfo {author} {\bibfnamefont {C.~H.}\ \bibnamefont
  {Monken}},\ and\ \bibinfo {author} {\bibfnamefont {R.~L.}\ \bibnamefont
  {de~Matos~Filho}},\ }\bibfield  {title} {\bibinfo {title} {Effects of spatial
  transverse correlations in second-harmonic generation},\ }\href
  {https://doi.org/10.1103/PhysRevA.73.053819} {\bibfield  {journal} {\bibinfo
  {journal} {Phys. Rev. A}\ }\textbf {\bibinfo {volume} {73}},\ \bibinfo
  {pages} {053819} (\bibinfo {year} {2006})}\BibitemShut {NoStop}%
\bibitem [{\citenamefont {Zhang}\ \emph {et~al.}(2010)\citenamefont {Zhang},
  \citenamefont {Wen}, \citenamefont {Zhu},\ and\ \citenamefont
  {Xiao}}]{Zhang2010}%
  \BibitemOpen
  \bibfield  {author} {\bibinfo {author} {\bibfnamefont {Y.}~\bibnamefont
  {Zhang}}, \bibinfo {author} {\bibfnamefont {J.}~\bibnamefont {Wen}}, \bibinfo
  {author} {\bibfnamefont {S.~N.}\ \bibnamefont {Zhu}},\ and\ \bibinfo {author}
  {\bibfnamefont {M.}~\bibnamefont {Xiao}},\ }\bibfield  {title} {\bibinfo
  {title} {Nonlinear talbot effect},\ }\href
  {https://doi.org/10.1103/physrevlett.104.183901} {\bibfield  {journal}
  {\bibinfo  {journal} {Physical Review Letters}\ }\textbf {\bibinfo {volume}
  {104}},\ \bibinfo {pages} {183901} (\bibinfo {year} {2010})}\BibitemShut
  {NoStop}%
\bibitem [{\citenamefont {Hong}\ \emph {et~al.}(2014)\citenamefont {Hong},
  \citenamefont {Yang}, \citenamefont {Zhang}, \citenamefont {Qin},\ and\
  \citenamefont {Zhu}}]{Hong2014}%
  \BibitemOpen
  \bibfield  {author} {\bibinfo {author} {\bibfnamefont {X.-H.}\ \bibnamefont
  {Hong}}, \bibinfo {author} {\bibfnamefont {B.}~\bibnamefont {Yang}}, \bibinfo
  {author} {\bibfnamefont {C.}~\bibnamefont {Zhang}}, \bibinfo {author}
  {\bibfnamefont {Y.-Q.}\ \bibnamefont {Qin}},\ and\ \bibinfo {author}
  {\bibfnamefont {Y.-Y.}\ \bibnamefont {Zhu}},\ }\bibfield  {title} {\bibinfo
  {title} {Nonlinear volume holography for wave-front engineering},\ }\href
  {https://doi.org/10.1103/physrevlett.113.163902} {\bibfield  {journal}
  {\bibinfo  {journal} {Physical Review Letters}\ }\textbf {\bibinfo {volume}
  {113}},\ \bibinfo {pages} {163902} (\bibinfo {year} {2014})}\BibitemShut
  {NoStop}%
\bibitem [{\citenamefont {Liu}\ \emph {et~al.}(2018)\citenamefont {Liu},
  \citenamefont {Zhao}, \citenamefont {Li}, \citenamefont {Zheng},\ and\
  \citenamefont {Chen}}]{Liu2018}%
  \BibitemOpen
  \bibfield  {author} {\bibinfo {author} {\bibfnamefont {H.}~\bibnamefont
  {Liu}}, \bibinfo {author} {\bibfnamefont {X.}~\bibnamefont {Zhao}}, \bibinfo
  {author} {\bibfnamefont {H.}~\bibnamefont {Li}}, \bibinfo {author}
  {\bibfnamefont {Y.}~\bibnamefont {Zheng}},\ and\ \bibinfo {author}
  {\bibfnamefont {X.}~\bibnamefont {Chen}},\ }\bibfield  {title} {\bibinfo
  {title} {Dynamic computer-generated nonlinear optical holograms in a
  non-collinear second-harmonic generation process},\ }\href
  {https://doi.org/10.1364/ol.43.003236} {\bibfield  {journal} {\bibinfo
  {journal} {Optics Letters}\ }\textbf {\bibinfo {volume} {43}},\ \bibinfo
  {pages} {3236} (\bibinfo {year} {2018})}\BibitemShut {NoStop}%
\bibitem [{\citenamefont {Steinlechner}\ \emph {et~al.}(2016)\citenamefont
  {Steinlechner}, \citenamefont {Hermosa}, \citenamefont {Pruneri},\ and\
  \citenamefont {Torres}}]{Stein2016}%
  \BibitemOpen
  \bibfield  {author} {\bibinfo {author} {\bibfnamefont {F.}~\bibnamefont
  {Steinlechner}}, \bibinfo {author} {\bibfnamefont {N.}~\bibnamefont
  {Hermosa}}, \bibinfo {author} {\bibfnamefont {V.}~\bibnamefont {Pruneri}},\
  and\ \bibinfo {author} {\bibfnamefont {J.~P.}\ \bibnamefont {Torres}},\
  }\bibfield  {title} {\bibinfo {title} {Frequency conversion of structured
  light},\ }\href {https://doi.org/10.1038/srep21390} {\bibfield  {journal}
  {\bibinfo  {journal} {Scientific Reports}\ }\textbf {\bibinfo {volume} {6}},\
  \bibinfo {pages} {21390} (\bibinfo {year} {2016})}\BibitemShut {NoStop}%
\bibitem [{\citenamefont {Wu}\ \emph {et~al.}(2022)\citenamefont {Wu},
  \citenamefont {Yu}, \citenamefont {Zhu}, \citenamefont {Gao}, \citenamefont
  {Ding}, \citenamefont {Zhou}, \citenamefont {Hu}, \citenamefont
  {Rosales-Guzm\'{a}n}, \citenamefont {Shen},\ and\ \citenamefont
  {Shi}}]{wu22}%
  \BibitemOpen
  \bibfield  {author} {\bibinfo {author} {\bibfnamefont {H.-J.}\ \bibnamefont
  {Wu}}, \bibinfo {author} {\bibfnamefont {B.-S.}\ \bibnamefont {Yu}}, \bibinfo
  {author} {\bibfnamefont {Z.-H.}\ \bibnamefont {Zhu}}, \bibinfo {author}
  {\bibfnamefont {W.}~\bibnamefont {Gao}}, \bibinfo {author} {\bibfnamefont
  {D.-S.}\ \bibnamefont {Ding}}, \bibinfo {author} {\bibfnamefont {Z.-Y.}\
  \bibnamefont {Zhou}}, \bibinfo {author} {\bibfnamefont {X.-P.}\ \bibnamefont
  {Hu}}, \bibinfo {author} {\bibfnamefont {C.}~\bibnamefont
  {Rosales-Guzm\'{a}n}}, \bibinfo {author} {\bibfnamefont {Y.}~\bibnamefont
  {Shen}},\ and\ \bibinfo {author} {\bibfnamefont {B.-S.}\ \bibnamefont
  {Shi}},\ }\bibfield  {title} {\bibinfo {title} {Conformal frequency
  conversion for arbitrary vectorial structured light},\ }\href
  {https://doi.org/10.1364/OPTICA.444685} {\bibfield  {journal} {\bibinfo
  {journal} {Optica}\ }\textbf {\bibinfo {volume} {9}},\ \bibinfo {pages} {187}
  (\bibinfo {year} {2022})}\BibitemShut {NoStop}%
\bibitem [{\citenamefont {Buono}\ \emph {et~al.}(2014)\citenamefont {Buono},
  \citenamefont {Moraes}, \citenamefont {Huguenin}, \citenamefont {Souza},\
  and\ \citenamefont {Khoury}}]{buono14}%
  \BibitemOpen
  \bibfield  {author} {\bibinfo {author} {\bibfnamefont {W.~T.}\ \bibnamefont
  {Buono}}, \bibinfo {author} {\bibfnamefont {L.~F.~C.}\ \bibnamefont
  {Moraes}}, \bibinfo {author} {\bibfnamefont {J.~A.~O.}\ \bibnamefont
  {Huguenin}}, \bibinfo {author} {\bibfnamefont {C.~E.~R.}\ \bibnamefont
  {Souza}},\ and\ \bibinfo {author} {\bibfnamefont {A.~Z.}\ \bibnamefont
  {Khoury}},\ }\bibfield  {title} {\bibinfo {title} {Arbitrary orbital angular
  momentum addition in second harmonic generation},\ }\href
  {https://doi.org/10.1088/1367-2630/16/9/093041} {\bibfield  {journal}
  {\bibinfo  {journal} {New Journal of Physics}\ }\textbf {\bibinfo {volume}
  {16}},\ \bibinfo {pages} {093041} (\bibinfo {year} {2014})}\BibitemShut
  {NoStop}%
\bibitem [{\citenamefont {Pereira}\ \emph {et~al.}(2017)\citenamefont
  {Pereira}, \citenamefont {Buono}, \citenamefont {Tasca}, \citenamefont
  {Dechoum},\ and\ \citenamefont {Khoury}}]{buono17}%
  \BibitemOpen
  \bibfield  {author} {\bibinfo {author} {\bibfnamefont {L.~J.}\ \bibnamefont
  {Pereira}}, \bibinfo {author} {\bibfnamefont {W.~T.}\ \bibnamefont {Buono}},
  \bibinfo {author} {\bibfnamefont {D.~S.}\ \bibnamefont {Tasca}}, \bibinfo
  {author} {\bibfnamefont {K.}~\bibnamefont {Dechoum}},\ and\ \bibinfo {author}
  {\bibfnamefont {A.~Z.}\ \bibnamefont {Khoury}},\ }\bibfield  {title}
  {\bibinfo {title} {Orbital-angular-momentum mixing in type-ii second-harmonic
  generation},\ }\href {https://doi.org/10.1103/PhysRevA.96.053856} {\bibfield
  {journal} {\bibinfo  {journal} {Phys. Rev. A}\ }\textbf {\bibinfo {volume}
  {96}},\ \bibinfo {pages} {053856} (\bibinfo {year} {2017})}\BibitemShut
  {NoStop}%
\bibitem [{\citenamefont {Buono}\ \emph {et~al.}(2018)\citenamefont {Buono},
  \citenamefont {Santiago}, \citenamefont {Pereira}, \citenamefont {Tasca},
  \citenamefont {Dechoum},\ and\ \citenamefont {Khoury}}]{buono18}%
  \BibitemOpen
  \bibfield  {author} {\bibinfo {author} {\bibfnamefont {W.~T.}\ \bibnamefont
  {Buono}}, \bibinfo {author} {\bibfnamefont {J.}~\bibnamefont {Santiago}},
  \bibinfo {author} {\bibfnamefont {L.~J.}\ \bibnamefont {Pereira}}, \bibinfo
  {author} {\bibfnamefont {D.~S.}\ \bibnamefont {Tasca}}, \bibinfo {author}
  {\bibfnamefont {K.}~\bibnamefont {Dechoum}},\ and\ \bibinfo {author}
  {\bibfnamefont {A.~Z.}\ \bibnamefont {Khoury}},\ }\bibfield  {title}
  {\bibinfo {title} {Polarization-controlled orbital angular momentum switching
  in nonlinear wave mixing},\ }\href {https://doi.org/10.1364/OL.43.001439}
  {\bibfield  {journal} {\bibinfo  {journal} {Opt. Lett.}\ }\textbf {\bibinfo
  {volume} {43}},\ \bibinfo {pages} {1439} (\bibinfo {year}
  {2018})}\BibitemShut {NoStop}%
\bibitem [{\citenamefont {Buono}\ \emph {et~al.}(2020)\citenamefont {Buono},
  \citenamefont {Santos}, \citenamefont {Maia}, \citenamefont {Pereira},
  \citenamefont {Tasca}, \citenamefont {Dechoum}, \citenamefont {Ruchon},\ and\
  \citenamefont {Khoury}}]{buono20}%
  \BibitemOpen
  \bibfield  {author} {\bibinfo {author} {\bibfnamefont {W.~T.}\ \bibnamefont
  {Buono}}, \bibinfo {author} {\bibfnamefont {A.}~\bibnamefont {Santos}},
  \bibinfo {author} {\bibfnamefont {M.~R.}\ \bibnamefont {Maia}}, \bibinfo
  {author} {\bibfnamefont {L.~J.}\ \bibnamefont {Pereira}}, \bibinfo {author}
  {\bibfnamefont {D.~S.}\ \bibnamefont {Tasca}}, \bibinfo {author}
  {\bibfnamefont {K.}~\bibnamefont {Dechoum}}, \bibinfo {author} {\bibfnamefont
  {T.}~\bibnamefont {Ruchon}},\ and\ \bibinfo {author} {\bibfnamefont {A.~Z.}\
  \bibnamefont {Khoury}},\ }\bibfield  {title} {\bibinfo {title} {Chiral
  relations and radial-angular coupling in nonlinear interactions of optical
  vortices},\ }\href {https://doi.org/10.1103/PhysRevA.101.043821} {\bibfield
  {journal} {\bibinfo  {journal} {Phys. Rev. A}\ }\textbf {\bibinfo {volume}
  {101}},\ \bibinfo {pages} {043821} (\bibinfo {year} {2020})}\BibitemShut
  {NoStop}%
\bibitem [{\citenamefont {de~Oliveira}\ \emph {et~al.}(2021)\citenamefont
  {de~Oliveira}, \citenamefont {Santos}, \citenamefont {da~Silva},
  \citenamefont {Pereira}, \citenamefont {Alves}, \citenamefont {Khoury},\ and\
  \citenamefont {Ribeiro}}]{Andre21}%
  \BibitemOpen
  \bibfield  {author} {\bibinfo {author} {\bibfnamefont {A.}~\bibnamefont
  {de~Oliveira}}, \bibinfo {author} {\bibfnamefont {G.}~\bibnamefont {Santos}},
  \bibinfo {author} {\bibfnamefont {N.~R.}\ \bibnamefont {da~Silva}}, \bibinfo
  {author} {\bibfnamefont {L.}~\bibnamefont {Pereira}}, \bibinfo {author}
  {\bibfnamefont {G.}~\bibnamefont {Alves}}, \bibinfo {author} {\bibfnamefont
  {A.}~\bibnamefont {Khoury}},\ and\ \bibinfo {author} {\bibfnamefont {P.~S.}\
  \bibnamefont {Ribeiro}},\ }\bibfield  {title} {\bibinfo {title} {Beyond
  conservation of orbital angular momentum in stimulated parametric
  down-conversion},\ }\href {https://doi.org/10.1103/PhysRevApplied.16.044019}
  {\bibfield  {journal} {\bibinfo  {journal} {Phys. Rev. Applied}\ }\textbf
  {\bibinfo {volume} {16}},\ \bibinfo {pages} {044019} (\bibinfo {year}
  {2021})}\BibitemShut {NoStop}%
\bibitem [{\citenamefont {Hickmann}\ \emph {et~al.}(2010)\citenamefont
  {Hickmann}, \citenamefont {Fonseca}, \citenamefont {Soares},\ and\
  \citenamefont {Ch\'avez-Cerda}}]{Will10}%
  \BibitemOpen
  \bibfield  {author} {\bibinfo {author} {\bibfnamefont {J.~M.}\ \bibnamefont
  {Hickmann}}, \bibinfo {author} {\bibfnamefont {E.~J.~S.}\ \bibnamefont
  {Fonseca}}, \bibinfo {author} {\bibfnamefont {W.~C.}\ \bibnamefont
  {Soares}},\ and\ \bibinfo {author} {\bibfnamefont {S.}~\bibnamefont
  {Ch\'avez-Cerda}},\ }\bibfield  {title} {\bibinfo {title} {Unveiling a
  truncated optical lattice associated with a triangular aperture using light's
  orbital angular momentum},\ }\href
  {https://doi.org/10.1103/PhysRevLett.105.053904} {\bibfield  {journal}
  {\bibinfo  {journal} {Phys. Rev. Lett.}\ }\textbf {\bibinfo {volume} {105}},\
  \bibinfo {pages} {053904} (\bibinfo {year} {2010})}\BibitemShut {NoStop}%
\bibitem [{\citenamefont {Melo}\ \emph {et~al.}(2018)\citenamefont {Melo},
  \citenamefont {Jesus-Silva}, \citenamefont {Ch{\'a}vez-Cerda}, \citenamefont
  {Ribeiro},\ and\ \citenamefont {Soares}}]{Will18}%
  \BibitemOpen
  \bibfield  {author} {\bibinfo {author} {\bibfnamefont {L.~A.}\ \bibnamefont
  {Melo}}, \bibinfo {author} {\bibfnamefont {A.~J.}\ \bibnamefont
  {Jesus-Silva}}, \bibinfo {author} {\bibfnamefont {S.}~\bibnamefont
  {Ch{\'a}vez-Cerda}}, \bibinfo {author} {\bibfnamefont {P.~H.~S.}\
  \bibnamefont {Ribeiro}},\ and\ \bibinfo {author} {\bibfnamefont {W.~C.}\
  \bibnamefont {Soares}},\ }\bibfield  {title} {\bibinfo {title} {Direct
  measurement of the topological charge in elliptical beams using diffraction
  by a triangular aperture},\ }\href
  {https://doi.org/10.1038/s41598-018-24928-5} {\bibfield  {journal} {\bibinfo
  {journal} {Scientific Reports}\ }\textbf {\bibinfo {volume} {8}},\ \bibinfo
  {pages} {6370} (\bibinfo {year} {2018})}\BibitemShut {NoStop}%
\bibitem [{\citenamefont {Shen}\ \emph {et~al.}(2018)\citenamefont {Shen},
  \citenamefont {Fu},\ and\ \citenamefont {Gong}}]{Shen18}%
  \BibitemOpen
  \bibfield  {author} {\bibinfo {author} {\bibfnamefont {Y.}~\bibnamefont
  {Shen}}, \bibinfo {author} {\bibfnamefont {X.}~\bibnamefont {Fu}},\ and\
  \bibinfo {author} {\bibfnamefont {M.}~\bibnamefont {Gong}},\ }\bibfield
  {title} {\bibinfo {title} {Truncated triangular diffraction lattices and
  orbital-angular-momentum detection of vortex su(2) geometric modes},\ }\href
  {https://doi.org/10.1364/OE.26.025545} {\bibfield  {journal} {\bibinfo
  {journal} {Opt. Express}\ }\textbf {\bibinfo {volume} {26}},\ \bibinfo
  {pages} {25545} (\bibinfo {year} {2018})}\BibitemShut {NoStop}%
\bibitem [{\citenamefont {de~Oliveira}\ \emph {et~al.}(2019)\citenamefont
  {de~Oliveira}, \citenamefont {Arruda}, \citenamefont {Soares}, \citenamefont
  {Walborn}, \citenamefont {Khoury}, \citenamefont {Kanaan}, \citenamefont
  {Ribeiro},\ and\ \citenamefont {de~Ara{\'u}jo}}]{Oliveira19}%
  \BibitemOpen
  \bibfield  {author} {\bibinfo {author} {\bibfnamefont {A.~G.}\ \bibnamefont
  {de~Oliveira}}, \bibinfo {author} {\bibfnamefont {M.~F.~Z.}\ \bibnamefont
  {Arruda}}, \bibinfo {author} {\bibfnamefont {W.~C.}\ \bibnamefont {Soares}},
  \bibinfo {author} {\bibfnamefont {S.~P.}\ \bibnamefont {Walborn}}, \bibinfo
  {author} {\bibfnamefont {A.~Z.}\ \bibnamefont {Khoury}}, \bibinfo {author}
  {\bibfnamefont {A.}~\bibnamefont {Kanaan}}, \bibinfo {author} {\bibfnamefont
  {P.~H.~S.}\ \bibnamefont {Ribeiro}},\ and\ \bibinfo {author} {\bibfnamefont
  {R.~M.}\ \bibnamefont {de~Ara{\'u}jo}},\ }\bibfield  {title} {\bibinfo
  {title} {Phase conjugation and mode conversion in stimulated parametric
  down-conversion with orbital angular momentum: a geometrical
  interpretation},\ }\href {https://doi.org/10.1007/s13538-018-0614-4}
  {\bibfield  {journal} {\bibinfo  {journal} {Brazilian Journal of Physics}\
  }\textbf {\bibinfo {volume} {49}},\ \bibinfo {pages} {10} (\bibinfo {year}
  {2019})}\BibitemShut {NoStop}%
\bibitem [{\citenamefont {Zhang}\ \emph {et~al.}(2017)\citenamefont {Zhang},
  \citenamefont {Yu}, \citenamefont {Wu},\ and\ \citenamefont
  {Halasyamani}}]{Zhang2017}%
  \BibitemOpen
  \bibfield  {author} {\bibinfo {author} {\bibfnamefont {W.}~\bibnamefont
  {Zhang}}, \bibinfo {author} {\bibfnamefont {H.}~\bibnamefont {Yu}}, \bibinfo
  {author} {\bibfnamefont {H.}~\bibnamefont {Wu}},\ and\ \bibinfo {author}
  {\bibfnamefont {P.~S.}\ \bibnamefont {Halasyamani}},\ }\bibfield  {title}
  {\bibinfo {title} {Phase-matching in nonlinear optical compounds: A materials
  perspective},\ }\href {https://doi.org/10.1021/acs.chemmater.7b00243}
  {\bibfield  {journal} {\bibinfo  {journal} {Chemistry of Materials}\ }\textbf
  {\bibinfo {volume} {29}},\ \bibinfo {pages} {2655} (\bibinfo {year}
  {2017})}\BibitemShut {NoStop}%
\bibitem [{\citenamefont {Bloembergen}(1977)}]{bloembergen1977}%
  \BibitemOpen
  \bibfield  {author} {\bibinfo {author} {\bibfnamefont {N.}~\bibnamefont
  {Bloembergen}},\ }\href@noop {} {\emph {\bibinfo {title} {Nonlinear
  {{Optics}}}}}\ (\bibinfo  {publisher} {{Pearson Addison-Wesley}},\ \bibinfo
  {address} {{New York}},\ \bibinfo {year} {1977})\BibitemShut {NoStop}%
\bibitem [{\citenamefont {Bai}\ \emph {et~al.}(2019)\citenamefont {Bai},
  \citenamefont {Zhang}, \citenamefont {Wang}, \citenamefont {Fu},
  \citenamefont {Shao}, \citenamefont {Li}, \citenamefont {Wan}, \citenamefont
  {Li}, \citenamefont {Cao}, \citenamefont {Guo},\ and\ \citenamefont
  {Shen}}]{bai2019}%
  \BibitemOpen
  \bibfield  {author} {\bibinfo {author} {\bibfnamefont {P.}~\bibnamefont
  {Bai}}, \bibinfo {author} {\bibfnamefont {Y.}~\bibnamefont {Zhang}}, \bibinfo
  {author} {\bibfnamefont {T.}~\bibnamefont {Wang}}, \bibinfo {author}
  {\bibfnamefont {Z.}~\bibnamefont {Fu}}, \bibinfo {author} {\bibfnamefont
  {D.}~\bibnamefont {Shao}}, \bibinfo {author} {\bibfnamefont {Z.}~\bibnamefont
  {Li}}, \bibinfo {author} {\bibfnamefont {W.}~\bibnamefont {Wan}}, \bibinfo
  {author} {\bibfnamefont {H.}~\bibnamefont {Li}}, \bibinfo {author}
  {\bibfnamefont {J.}~\bibnamefont {Cao}}, \bibinfo {author} {\bibfnamefont
  {X.}~\bibnamefont {Guo}},\ and\ \bibinfo {author} {\bibfnamefont
  {W.}~\bibnamefont {Shen}},\ }\bibfield  {title} {\bibinfo {title} {Broadband
  {{THz}} to {{NIR}} up-converter for photon-type {{THz}} imaging},\ }\href
  {https://doi.org/10.1038/s41467-019-11465-6} {\bibfield  {journal} {\bibinfo
  {journal} {Nat Commun}\ }\textbf {\bibinfo {volume} {10}},\ \bibinfo {pages}
  {3513} (\bibinfo {year} {2019})}\BibitemShut {NoStop}%
\bibitem [{\citenamefont {Haase}\ \emph {et~al.}(2019)\citenamefont {Haase},
  \citenamefont {Kutas}, \citenamefont {Riexinger}, \citenamefont {Bickert},
  \citenamefont {Keil}, \citenamefont {Molter}, \citenamefont {Bortz},\ and\
  \citenamefont {von Freymann}}]{haase2019}%
  \BibitemOpen
  \bibfield  {author} {\bibinfo {author} {\bibfnamefont {B.}~\bibnamefont
  {Haase}}, \bibinfo {author} {\bibfnamefont {M.}~\bibnamefont {Kutas}},
  \bibinfo {author} {\bibfnamefont {F.}~\bibnamefont {Riexinger}}, \bibinfo
  {author} {\bibfnamefont {P.}~\bibnamefont {Bickert}}, \bibinfo {author}
  {\bibfnamefont {A.}~\bibnamefont {Keil}}, \bibinfo {author} {\bibfnamefont
  {D.}~\bibnamefont {Molter}}, \bibinfo {author} {\bibfnamefont
  {M.}~\bibnamefont {Bortz}},\ and\ \bibinfo {author} {\bibfnamefont
  {G.}~\bibnamefont {von Freymann}},\ }\bibfield  {title} {\bibinfo {title}
  {Spontaneous parametric down-conversion of photons at 660 nm to the terahertz
  and sub-terahertz frequency range},\ }\href
  {https://doi.org/10.1364/OE.27.007458} {\bibfield  {journal} {\bibinfo
  {journal} {Opt. Express}\ }\textbf {\bibinfo {volume} {27}},\ \bibinfo
  {pages} {7458} (\bibinfo {year} {2019})}\BibitemShut {NoStop}%
\bibitem [{\citenamefont {Jana}\ \emph {et~al.}(2021)\citenamefont {Jana},
  \citenamefont {Okocha}, \citenamefont {Møller}, \citenamefont {Mi},
  \citenamefont {Sederberg},\ and\ \citenamefont {Corkum}}]{jana2021}%
  \BibitemOpen
  \bibfield  {author} {\bibinfo {author} {\bibfnamefont {K.}~\bibnamefont
  {Jana}}, \bibinfo {author} {\bibfnamefont {E.}~\bibnamefont {Okocha}},
  \bibinfo {author} {\bibfnamefont {S.~H.}\ \bibnamefont {Møller}}, \bibinfo
  {author} {\bibfnamefont {Y.}~\bibnamefont {Mi}}, \bibinfo {author}
  {\bibfnamefont {S.}~\bibnamefont {Sederberg}},\ and\ \bibinfo {author}
  {\bibfnamefont {P.~B.}\ \bibnamefont {Corkum}},\ }\bibfield  {title}
  {\bibinfo {title} {Reconfigurable terahertz metasurfaces coherently
  controlled by wavelength-scale-structured light},\ }\bibfield  {journal}
  {\bibinfo  {journal} {Nanophotonics}\ }\href
  {https://doi.org/doi:10.1515/nanoph-2021-0501} {doi:10.1515/nanoph-2021-0501}
  (\bibinfo {year} {2021})\BibitemShut {NoStop}%
\bibitem [{\citenamefont {Rocha}\ \emph {et~al.}(2021)\citenamefont {Rocha},
  \citenamefont {Pires}, \citenamefont {Neto}, \citenamefont {Jesus-Silva},
  \citenamefont {Litchinitser},\ and\ \citenamefont {Fonseca}}]{rocha2021}%
  \BibitemOpen
  \bibfield  {author} {\bibinfo {author} {\bibfnamefont {J.~C.~A.}\
  \bibnamefont {Rocha}}, \bibinfo {author} {\bibfnamefont {D.~G.}\ \bibnamefont
  {Pires}}, \bibinfo {author} {\bibfnamefont {J.~G. M.~N.}\ \bibnamefont
  {Neto}}, \bibinfo {author} {\bibfnamefont {A.~J.}\ \bibnamefont
  {Jesus-Silva}}, \bibinfo {author} {\bibfnamefont {N.~M.}\ \bibnamefont
  {Litchinitser}},\ and\ \bibinfo {author} {\bibfnamefont {E.~J.~S.}\
  \bibnamefont {Fonseca}},\ }\bibfield  {title} {\bibinfo {title} {Speckle
  filtering through nonlinear wave mixing},\ }\href
  {https://doi.org/10.1364/OL.434150} {\bibfield  {journal} {\bibinfo
  {journal} {Opt. Lett.}\ }\textbf {\bibinfo {volume} {46}},\ \bibinfo {pages}
  {3905} (\bibinfo {year} {2021})}\BibitemShut {NoStop}%
\bibitem [{\citenamefont {Castelvecchi}(2021)}]{Castelvecchi21}%
  \BibitemOpen
  \bibfield  {author} {\bibinfo {author} {\bibfnamefont {D.}~\bibnamefont
  {Castelvecchi}},\ }\bibfield  {title} {\bibinfo {title} {Quantum network is
  step towards ultrasecure internet},\ }\href
  {https://doi.org/10.1038/d41586-021-00420-5} {\bibfield  {journal} {\bibinfo
  {journal} {Nature}\ }\textbf {\bibinfo {volume} {590}},\ \bibinfo {pages}
  {540} (\bibinfo {year} {2021})}\BibitemShut {NoStop}%
\end{thebibliography}%
%\clearpage
%\appendix
%\section{...}

\end{document}